\documentclass{mn2e}
\usepackage{graphicx}
\usepackage{epsfig}
\usepackage{amsmath}
\usepackage[T1]{fontenc}
\usepackage{times}
\newcommand{\gae}
{\lower 2pt \hbox{$\, \buildrel {\scriptstyle >}\over {\scriptstyle \sim}\,$}}
\newcommand{\lae}
{\lower 2pt \hbox{$\, \buildrel {\scriptstyle <}\over {\scriptstyle \sim}\,$}}

\newcommand{\0}{\phantom{0}}

\begin{document}

\title[Meteor showers on Earth from sungrazing comets]
{Meteor showers on Earth from sungrazing comets}

\author[A. Sekhar and D. J. Asher]
{A. Sekhar$^{1,2*}$ and D. J. Asher$^1$\\
 $^1$Armagh Observatory, College Hill, Armagh BT61\ 9DG\\
 $^2$Queen's University of Belfast, University Road, Belfast BT7 1NN\\
 $^*$E-mail: asw@arm.ac.uk , asekhar01@qub.ac.uk \\ }

\date{{\bf Accepted}: 2013 Oct 11; {\bf Received}: 2013 Oct 3; {\bf In Original Form}: 2013 Sep 20; {\bf Accepted by MNRAS Letters}}

\maketitle

\begin{abstract}
Sungrazing comets have always captured a lot of interest and curiosity among
the general public as well as scientists since ancient times. The perihelion
passage of comet C/2012 S1 (ISON) at the end of this year (on 2013 November 28) is
an eagerly awaited event. In this work, we do a mathematical study to check
whether meteoroids ejected from this comet during its journey around the sun
can produce spectacular meteor phenomena on Earth. Our calculations show that
although the orbital elements of this comet are much more favourable than for
most sungrazers to have its descending node near the Earth's orbit, even
ejection velocities as high as 1 km s$^{-1}$ do not induce sufficient nodal
dispersion to bring meteoroids to Earth intersection during present times.
A similar result applies to Newton's comet C/1680 V1 which has
surprisingly similar orbital elements, although it is known to be a distinct
comet from C/2012 S1.
Our analysis also shows that for meteoroids ejected
from all known sungrazing groups during recent epochs, only the Marsden
family (with required ejection velocities of some hundreds of m s$^{-1}$) can
produce meteor phenomena during present times. In a broader sense, we
indicate why we do not observe visually brilliant meteor showers from
frequently observed sungrazers.

\end{abstract}

\begin{keywords}
celestial mechanics; meteoroids, meteors, meteorites; comets:
  general; comets: individual: C/2012 S1 (ISON), C/1680 V1 (Newton's comet)

\end{keywords}

\section{Introduction}

There have been various interesting observational records (Marsden 1967,
1989, 2005; Strom 2002; Sekanina \& Chodas 2012) of extremely bright and
spectacular sungrazing comets since historical times. Intuitively one would
expect some of these to produce spectacular meteor showers (like those
from many Jupiter family and Halley type comets). In case of such a shower
when sungrazers are involved, there are two factors in favour of producing
intense meteor phenomena. Firstly these comets pass very close to the sun ($q
\sim 0.004-0.06 $ au) which would enable more ices to sublime according to
conventional understanding (Whipple 1950) and thereby eject more dust
particles into similar orbits. Secondly some sungrazers are dynamically new
comets (Bailey, Chambers \& Hahn 1992), coming from the Oort-\"Opik cloud
into the inner solar system for the first time, suggesting a strong
possibility for more volatiles in their composition (enhancing chances for
strong outgassing).

Nevertheless we hardly observe any spectacular meteor activity on Earth due
to these frequently observed sungrazing comets. This work presents a
mathematical formalism of demonstrating the absence of any strong meteor
shower from comet C/2012 S1 (ISON). During this analysis some parallels are
drawn with the famous Newton's comet C/1680 V1 due to the surprisingly similar orbital elements. The same technique is then applied to all the known
sungrazing families.

\section{Effect of ejection velocity on meteoroids' nodal distances}
\label{}

We use the notation:\\
$a$ (semi-major axis),
$e$ (eccentricity),
$q$ (perihelion distance),
$i$ (inclination),
$\omega$ (argument of pericentre),
$\Omega$ (longitude of ascending node),
$\varpi$ (longitude of pericentre),
$E$ (eccentric anomaly),
$f$ (true anomaly),
$S, dv_{r}$ (radial component of meteoroid ejection acceleration/velocity),
$T, dv_{t}$ (transverse component = in-plane, orthogonal to radial),
$W, dv_{n}$ (normal component),
$n$ (mean motion),
$G$ (universal gravitational constant),
$M$ (mass of sun),
$t$ (time),
$r$ (heliocentric distance),
$r_{a}, r_{d}$ (heliocentric distance of ascending/descending node)

\subsection{Conditions to favour meteor phenomena on Earth}
Among the most critical parameters determining the feasibility of meteor
showers on Earth are the ascending and descending nodal distances of meteoroid
particles:

\begin{equation}
r_a = \frac{a(1-e^{2})}{(1+e \cos \omega)} = \frac{q(1+e)}{(1+e \cos \omega)} 
\end{equation}

 \vspace*{-4mm}

\begin{equation}
r_d = \frac{a(1-e^{2})}{(1-e \cos \omega)} = \frac{q(1+e)}{(1-e \cos \omega)}
\end{equation}

The necessary condition (but not sufficient) for meteor activity on Earth is:
$r_a \sim 1$ au or $r_d \sim 1$ au. This implies

\begin{equation}
\omega=\cos^{-1}[(q(1+e)-1)/e]
\end{equation}

 \vspace*{-7mm}

\begin{equation}
\omega=\cos^{-1}[(1-q(1+e))/e]
\end{equation}

From the compiled observations of sungrazers (Marsden \& Williams 2008) one
can constrain the range of $e$ and $q$. The condition $e \sim 1$ simplifies
equations (3) and (4) to

\begin{equation}
\omega = \cos^{-1}[2q-1]
\end{equation}

 \vspace*{-7mm}

\begin{equation}
\omega = \cos^{-1}[1-2q]
\end{equation}

For the range $q \sim$ [0.004 au, 0.06 au] which comes from observations of
various sungrazing families: equation (5) shows $r_a \sim 1$ au only if $\omega \sim$ [152,173], [187,208]; equation (6) shows $r_d \sim 1$ au only if $\omega \sim$ [7,28], [332,353]. Each interval spans $\sim$21\degr, and $\omega$ in one of these four ranges
is a necessary (but not sufficient) condition for high $e$ sungrazers to
undergo meteoroid intersection with Earth.

Although confirming the presence of a meteor shower on Earth would depend on
other parameters like time of nodal crossing, Earth's precise position in its
own orbit at that time and width of the dust trail, confirming the absence of
significant meteor activity can be done using this necessary condition
concerning the geometry of nodes.

\subsection{Separating the effects due to three components of ejection velocity}

Even if parent bodies' nodal distances are quite far from Earth's orbit,
meteoroid ejection in different directions can change the nodal distances
into $r_a \pm dr_a$ and $r_d \pm dr_d$ depending on the ejection velocity
components. Therefore checking these parameters for realistic values of
cometary ejection velocities can verify whether the meteoroid stream's nodal
distances can approach 1 au. 

The mathematical technique underlying our analysis uses Lagrange's planetary
equations:

\begin{equation}
\frac{da}{dt} = \frac{2}{n\sqrt{1-e^{2}}}(Se\sin f+\frac{a(1-e^{2})T}{r})
\end{equation}

\begin{equation}
\frac{de}{dt} = \frac{\sqrt{1-e^{2}}}{na}(S\sin f+T(\cos E+\cos f))
\end{equation}

\begin{equation}
\frac{d \varpi}{dt} = \frac{\sqrt{1-e^{2}}}{nae}[-S\cos f + T(1+\frac{r}{a(1-e^{2})})\sin f] + 2 \frac{d \Omega}{dt} \sin^{2}(\frac{i}{2})
\end{equation}

\begin{equation}
\frac{d \Omega}{dt} = \frac{Wr\sin (\omega +f)}{(na^{2}\sqrt{1-e^{2}}\sin i)}
\end{equation}
Equations (7) to (10) are taken from page 184, Roy (1978). Using the definition  
$\varpi \equiv \Omega + \omega$ we have:

\begin{equation}
\begin{split}
\frac{d \omega}{dt} =[ -\cos f \frac{\sqrt {1-e^{2}}}{nae}]\ S + [\sin f(1+\frac{r}{a(1-e^{2})})\frac{\sqrt {1-e^{2}}}{nae}] \ T & \\ + [(2\sin^{2}(\frac{i}{2})-1)(\frac{r\sin (\omega+f)}{na^{2}\sqrt{1-e^{2}}\ \sin i})]W
\end{split}
\end{equation}
Equation (11) can be shown to be equivalent to the expression on page
57, Murray \& Dermott (1999):

\begin{equation}
\begin{split}
\frac{d \omega}{dt} =  e^{-1} \sqrt{a \mu^{-1}  (1-e^{2})}[-S \cos f + T \sin f (\frac{2+e \cos f}{1+e \cos f})] & \\ -(\frac{d \Omega}{dt}) \cos i
\end{split}
\end{equation}
(where $\mu = GM$), which confirms that our substitutions from the
fundamental equations given by Roy (1978) yield the right result.

Taking the differential of equation (1) and finding the expressions for the partial derivatives gives


\begin{equation}
\begin{split}
dr_{a} = [\frac{(1-e^{2})}{1+e\cos\omega}]da + [\frac{-2ae(1+e\cos\omega)-a\cos\omega(1-e^{2})}{(1+e\cos\omega)^{2}}]de & \\ +[\frac{ae(1-e^{2})\sin\omega}{(1+e\cos\omega)^{2}}]d\omega
\end{split}
\end{equation}
Similarly equation (2) leads to


\begin{equation}
\begin{split}
dr_{d} = [\frac{(1-e^{2})}{1-e\cos\omega}]da + [\frac{-2ae(1-e\cos\omega)+a\cos\omega(1-e^{2})}{(1-e\cos\omega)^{2}}]de & \\ +  [\frac{-ae(1-e^{2})\sin\omega}{(1-e\cos\omega)^{2}}]d\omega
\end{split}
\end{equation}

Equations (13) and (14) require expressions for $da$, $de$ and $d \omega$.
These orbital element changes can be related to the separate velocity
components using (7), (8) and (11):

\begin{equation}
da =[\frac{2}{n\sqrt{1-e^{2}}}e\sin f]dv_{r}+[\frac{2a\sqrt{(1-e^{2})}}{nr}]dv_{t}+[0]dv_{n}
\end{equation}

\begin{equation}
de = [\frac{\sqrt{1-e^{2}}}{na}\sin f]dv_{r}+[\frac{\sqrt{1-e^{2}}}{na}(\cos E+\cos f)]dv_{t}+[0]dv_{n}
\end{equation}

\begin{equation}
\begin{split}
{d \omega} =[ -\cos f \frac{\sqrt {1-e^{2}}}{nae}]\ dv_{r} + [\sin f(1+\frac{r}{a(1-e^{2})})\frac{\sqrt {1-e^{2}}}{nae}] \ dv_{t} & \\ +[(2\sin^{2}(\frac{i}{2})-1)(\frac{r\sin (\omega+f)}{na^{2}\sqrt{1-e^{2}}\sin i})]dv_{n}
\end{split}
\end{equation}

Equations (15), (16), (17) followed by (13) or (14) express the changes in
ascending and descending nodal distances as linear combinations of the
separate radial, transverse and normal ejection velocity components at any
given point in the orbit. Numerical checks confirmed these differential
approximations to be good for the ranges in $dv_r$, $dv_t$, $dv_n$ up to
$\pm$1 km s$^{-1}$ where we apply them.

\section{C/2012 S1 (ISON) and C/1680 V1 (Newton's comet) }

\begin{figure}
\includegraphics[width=3in]{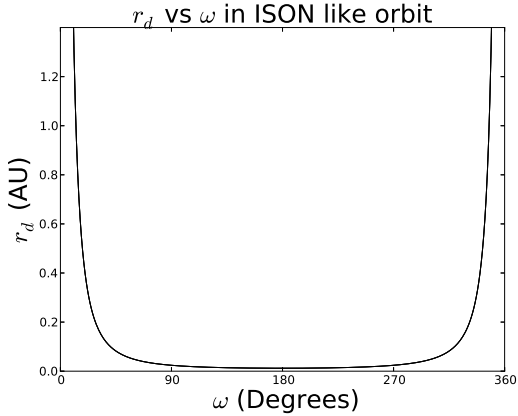}
\\[-1mm]
\caption{Heliocentric distances of descending node versus
  argument of pericentre for an ISON like ($q \sim$ 0.012 au, $e \sim $ 1)
  orbit: $r_d \sim$ 1 au when $\omega \sim$ 12 or 348\degr. The real comet ISON's $\omega \sim$
  346\degr\ is close to this value. }
\label{omega-fig}
\end{figure}

\begin{table}
\centering
\caption{Orbital elements taken from JPL Horizons (Giorgini et al.\ 1996),
  IAU Minor Planet Center, and computed nodal distances, for a few well known
  sungrazers namely ISON, Lovejoy, Ikeya-Seki \& Newton's comet respectively;
  $e \sim$ 1 for all of them.}
\label{orbels}
\begin{tabular}{@{}rccrrcc}
\hline
Comet                      &  $q$    &  $i$ &$\omega$&$r_a$& $r_d$ \\
                         & (au)  & (Degrees) & (Degrees) & (au)  & (au) \\
\hline
C/2012 S1            & 0.012  & \061.8 &345.536 & 0.012 &0.755 \\
C/2011 W3         & 0.006 &134.3 & \053.877 & 0.007 &0.029 \\
C/1965 S1-A    & 0.008  &141.9 & \069.049 & 0.012 &0.025 \\
C/1680 V1       & 0.006   &\060.5 &350.613 & 0.006 &0.881 \\
\hline
\end{tabular}\\
\end{table}

Comet C/2012 S1 (ISON) is predicted to pass very close ($q \sim 0.012$ au) to
the sun on 2013 November 28 (Samarasinha \& Mueller 2013;
Knight \& Walsh 2013) and hence expected to be a spectacular sungrazer this
year. ISON's $\omega \sim$ 346\degr\ $\in$ [332,353] lies in the favourable
range (Section 2.1) for which the descending node can be close to Earth's
orbit, i.e.\ the perihelion direction is quite favourably oriented for
potential meteor showers. Specifically, for an ISON like orbit ($q$=0.012 au
and $e \sim$ 1), equations (5) and (6) imply $r_a \sim 1$ au when $\omega
\sim$ 168 or 192\degr\ and $r_d \sim$ 1 au when $\omega \sim$ 12 or
348\degr\ (Fig.\ \ref{omega-fig}), and equation (2) shows comet ISON itself
has $r_d \sim 0.76$ au (see Table \ref{orbels}).

\begin{figure}
(a) \includegraphics[width=3in]{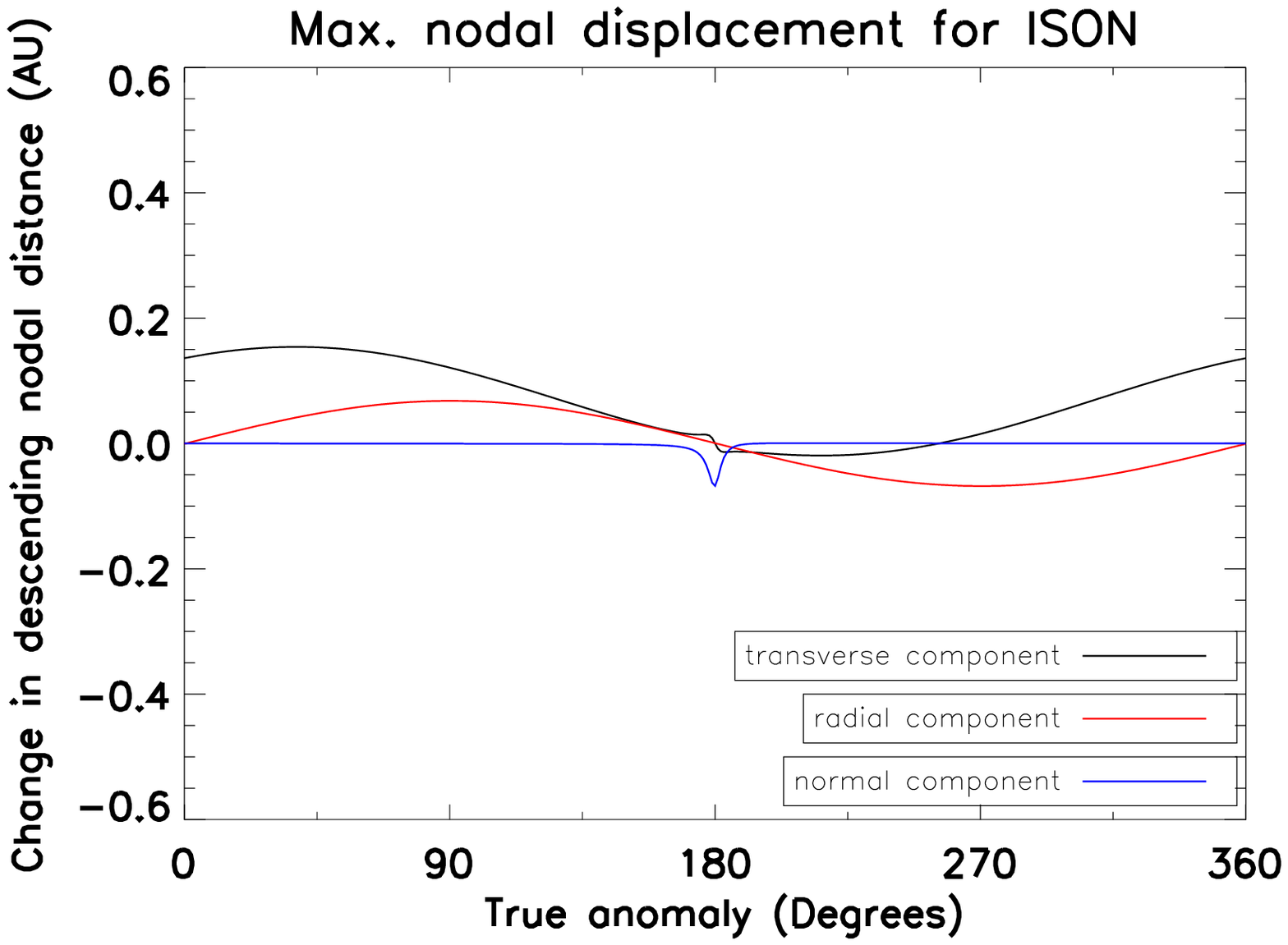}\\
(b) \includegraphics[width=3in]{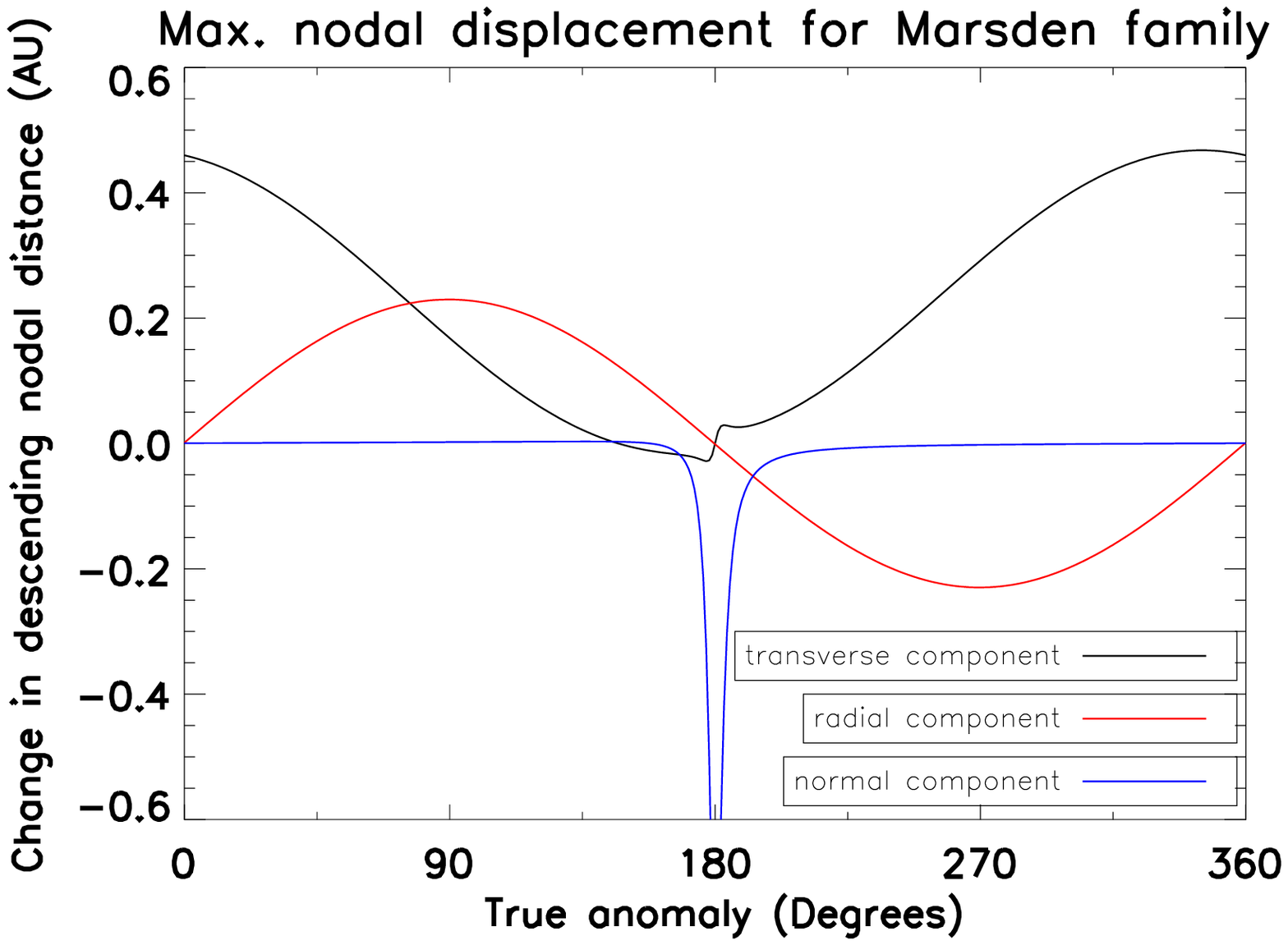}\\
\caption{Effect of each component of ejection velocity on descending nodal
  distance, as a function of true anomaly when transverse (black line); radial (red line); normal component (blue line) = 1 km s$^{-1}$ for (a) C/2012 S1 ISON (b) Marsden
  family comets.}
\label{drd-fig}
\end{figure}

Calculations outlined in Section 2.2 were done to find the nodal dispersion
of meteoroids ejected over the entire range of true anomaly from
0--360\degr. In reality, for water ice sublimation one needs to check only $r
\sim 0.012-3.4$ au (cf.\ Fitzsimmons \& Williams 1994) which corresponds to
$f \in [-174,174]$ here. Previous work on comet C/1995 O1 (Hale-Bopp) has
shown that CO outgassing could even occur around 7 au (Enzian 1999) which corresponds to $f \in [-176,176]$ in
the case of ISON. Fig.\ \ref{drd-fig}(a) shows the change to $r_d$ at all
$f$ for a fixed value (1 km s$^{-1}$) of each velocity component. For small
enough changes as considered here, $dr_d$ is proportional to ejection
velocity, and the nodal displacements simply alternate signs depending on
whether the velocity is positive or negative. Table \ref{max-dr} shows the
maximum absolute change in $r_d$ (also in $r_a$, not plotted in Fig.\
\ref{drd-fig}) over all $f$ when the magnitude of the ejection velocity
component is $\pm$1 km s$^{-1}$.

Fig.\ \ref{drd-fig} and Table \ref{max-dr} show that changes in nodal
distance due to radial and normal ejection velocity components are much less
effective than those due to the transverse part in bringing the node close to
Earth's orbit. However, a value of $dv_t$ = 1 km s$^{-1}$, even at the most
suitable $f$, still fails to bring the orbit to Earth intersection ($r_d +
dr_d \sim 0.91 $ au).

Ejection velocities well above 1 km s$^{-1}$ could bring the node close to
Earth. Most well known meteor showers (from Jupiter family and Halley type
comets) show prominent activity due to meteoroids with ejection velocities of
the order of 10 m s$^{-1}$. This has been confirmed from
numerous earlier works comparing the prediction and accurate observation of
meteor outbursts (Asher, Bailey \& Emel'yanenko 1999; Brown \& Arlt 2000; Ma \& Williams 2001; 
Rendtel 2007; Sekhar \& Asher 2013).  Because $q$ is much less for sungrazers
compared to other types of comet, meteoroid ejection with higher speeds than
a few tens of m s$^{-1}$ is definitely feasible, e.g.\ ejection speeds
approximately proportional to $r^{-1}$ for small $r$ (cf.\ Whipple 1951;
Jones 1995; Ma, Williams \& Chen 2002). However according to Whipple's model, ejection
velocities in the range of a few km s$^{-1}$ are unrealistic even for low $q$
sungrazers.

Moreover the number of particles with diameters $\ga$ 1 mm, which produce
visually spectacular showers, would be small and hence they will not lead to
any intense activity. Also the evolution of particles of diameter
significantly less than 1 mm with such high ejection velocities will be
dominated by other forces (Nesvorn\'y et al. 2011) such as radiation pressure
and Poynting-Robertson (Burns, Lamy \& Soter 1979). Hence studying the
geometry of nodal dispersion due
to ejection velocities would not work efficiently to verify such rare (and
almost visually unobservable) processes.

Surprisingly the orbital elements of this new comet ISON in 2013 and the
famous Newton's comet of 1680 have similar orbital elements
(Table \ref{orbels}). This has even led to speculations that both might
have a common ancestor, although ISON's original period of millions of yr
(given on IAU Minor Planet Center website) is accurate enough to exclude
their actually being the same comet.
The value of $\omega \sim$ 351\degr\ (for C/1680 V1) is also favourable
(Section 2.1), and $r_d \sim 0.88$ au. Similar calculations were done
(Section 2.2) to check the nodal dispersion of ejected meteoroids. As with
ISON, radial and normal components are less effective (Table \ref{max-dr}) in
pushing the nodes towards Earth. In principle, the transverse component of
ejection velocity can bring the descending node ($r_d + dr_d \sim$ 1 au) very
close to Earth's orbit at very high ejection velocities ($dv_t \sim$ 800 m s$^{-1}$). The absence of any known predictions or observations by Newton or
other scientists at that time or later about possibly related spectacular
meteor phenomena prevents any further conclusions here. (If there had been
any such credible observations, it could be evidence for high ejection speeds
$\sim$ 1 km s$^{-1}$, provided other conditions were satisfied such as
meteoroids and Earth reaching their mutual node at the same time).

For comparison, the well known sungrazers C/2011 W3 (Lovejoy) and C/1965 S1-A
(Ikeya-Seki) do not have $\omega$ in the favourable range discussed in
Section 2.1. As expected their nodes are far away (see Table \ref{orbels})
from Earth's orbit. A similar analysis in terms of the ejection velocity
components showed the nodes of meteoroids remain well below 1 au
even at very high ejection velocities ($dv_t, dv_r, dv_n \sim 1$ km s$^{-1}$). Hence meteor activity on Earth from particles ejected from these
comets in present times can also be completely ruled out.

\vspace*{-5mm}

\section{Marsden family versus other sungrazing families}

A search among 1440 compiled orbits of sungrazers belonging to various known
sungrazing families (Marsden \& Williams 2008) showed that only 27 of these
orbits have $\omega$ in the favourable ranges mentioned in Section 2.1 (Table
\ref{family}). It is reasonable to assume that out of this small number of
favourable parent bodies, a few of them might have fallen into the sun or got
tidally fragmented (like the Kreutz family discussed in Biesecker et
al.\ 2002) which thereby makes the number of possible candidates even
smaller. Many of these sungrazers have very small sizes (Iseli at
al.\ 2002).

\begin{table}
\centering
\caption{Maximum nodal displacement of meteoroids due to individual components of ejection velocity}
\label{max-dr}
\begin{tabular}{@{}c@{\,\,}c@{\,\,\,}c@{\,\,\,}ccc@{}}
\hline
Comet &$ dv_r $  & $dv_t$ & $dv_n$ &$\mid dr_a\mid$ & $\mid dr_d\mid$   \\
  & (km s$^{-1}$)            & (km s$^{-1}$)     & (km s$^{-1}$)      & $\times 10^{-3}$(au)        & $\times 10^{-1}$(au)   \\
\hline

C/2012 S1                     & 0           &0   &$\pm$1    & 3.167&0.677    \\
 
                 & 0           &$\pm$1   &0    &  4.274 & 1.503     \\
                   & $\pm$1           &0   &0    &  3.166 & 0.668     \\

C/1680 V1    & 0           &0   &$\pm$1    &  1.022 & 0.569  \\

& 0           &$\pm$1   &0    &  1.524  & 1.589     \\
& $\pm$1           &0   &0    &  1.137 & 0.569     \\
                   
\hline
\end{tabular}\\
\end{table}

\begin{table}
\centering
\caption{Distribution of sungrazing families from Catalogue of Cometary
  Orbits 2008 (50 of 1490 sungrazing comets listed have not been linked to
  any specific families).}
\label{family}
\begin{tabular}{@{}rccrrcc}
\hline
Family   &Number of comets  & Bodies favouring the range in $\omega$     \\
         &                  & so that $r_a \sim$ 1 au or $r_d \sim 1$ au \\
\hline
Kreutz   &    1277             &\00      \\
Meyer    &    \0\089             &\00         \\
Marsden  &   \0\032             &27          \\
Kracht 1 &   \0\031             &\00          \\
Kracht 2 &    \0\0\04             &\00       \\
Kracht 3 &    \0\0\02             &\00          \\
Anon 1   &    \0\0\03             &\00          \\
Anon 2   &    \0\0\02             &\00         \\
\hline
\end{tabular}\\
\end{table}

Our calculations (using equations 1 and 2) confirm that $r_a$ and $r_d$ are
significantly less than 1 au for all other sungrazing families. Thus only
Marsden family comets have conditions favourable to produce meteoroids that
can encounter Earth in present times (although comets from other families
could have favourable conditions, in terms of the right combination of
$\omega$ and $q$, to produce meteor phenomena during their distant past or
future).

Marsden family members have $r_d \in [0.16,4.76]$ au. The range for the 27
most favourable members (cf.\ Table \ref{family}) is $r_d \in [0.81,4.63]$
au. Fig.\ \ref{drd-fig}(b) shows the change to $r_d$ at all $f$ due to
each ejection velocity component; these plots are virtually identical for all
Marsden family members. Fig.\ \ref{drd-fig}(b) shows that the transverse
component $dv_t$ is most effective in changing the nodal distance so that it
can come near 1 au.  For values of $f$ where $dv_t$ is ineffective, both
$dv_r$ and $dv_n$ can be significant (Fig.\ \ref{drd-fig}(b)), although
$dv_n$ is most effective near aphelion where normal meteoroid ejection is not
expected.

Earlier works (Seargent 2002; Ohtsuka, Nakano \& Yoshikawa 2003; Sekanina \& Chodas 2005;
Jenniskens 2006; Jenniskens, Duckworth \& Grigsby 2012) have proposed that the Marsden
family could be linked to the daytime Arietids (171 ARI, IAU-MDC).  Our
calculations show that Marsden family members typically need ejection
velocities of at least a few hundred m s$^{-1}$ so that $r_d \pm dr_d \sim $
1 au. The number of large meteoroids (diameters $\ga$ 1 mm) having high
ejection velocities of several hundred m s$^{-1}$ (required to bring their
nodes close to Earth's orbit in this case) will be quite small (cf.\
discussion in section 3). This could be an explanation if the Zenithal Hourly
Rate of visually observed Arietids is indeed low, at about 1--2 meteors/hour
(Jenniskens et al.\ 2012).

Our analysis specifically shows that ejection speeds of some hundreds of m s$^{-1}$ from most Marsden family sungrazers can produce meteoroids whose
descending node is at 1 au. This accords with the proposed association with
171 ARI which has $\omega \sim$ 20--30\degr, similar to the Marsden family
(cf.\ equation 6). Long term evolution to induce a substantial $\omega$
separation is not required. Various  Marsden family members in the
dataset we used had perihelion passages during 1996--2008, a range that can
easily arise in the short term (even a single revolution). In the short term,
the nodal distances $r_d$ remain in the range resulting from the ejection
velocities.

However orbital changes due
e.g.\ to planetary perturbations would be substantial during long term
evolution, over which different points in the $\omega$ precession cycle may
be reached. Previous works (Ohtsuka et al.\ 2003; Sekanina \& Chodas 2005)
have found that the Kracht group ($\omega \sim 50$\degr) could be linked to
the Marsden family and 171 ARI during their long term evolution.
Sekanina \& Chodas (2005) identified a possible connection between the
Southern $\delta$ Aquariids (005 SDA) which have $\omega \sim 150$\degr\
(meteor shower at ascending node; equation 5) and the Marsden and Kracht
groups.

The nominal orbital periods of most Marsden sungrazers are very high ($\sim
10^{3}-10^{6} $ yr) because $e \sim$ 1. Hence we have not checked the orbital
evolution of meteoroids for subsequent revolutions. At the upper end of this
period range, long term future predictions for meteor showers from this
family would even require a completely independent analysis including other
effects due to perturbations from galactic tides or passing stars. The same
limitation applies for the long term evolution of meteoroids from ISON where
original $1/a \sim 9 \times 10^{-6}$ au$^{-1}$ (IAU-MPC) and Newton's comet
where $a \sim 444$ au (JPL Horizons) as well. The accuracy of conventional
long term predictions (usually applied to Jupiter family and Halley type
comets) of meteor showers without considering these additional effects will
be questionable when orbital periods are very high (which applies to most
sungrazers).

A similar analysis (as for ISON and Marsden family) was done on
the orbits of other sungrazing families (mentioned in Table \ref{family}).
Our calculations clearly show that no realistic ejection velocities in any
direction can bring the nodes of meteoroids close to Earth's orbit, $r_a \pm
dr_a$ and $r_d \pm dr_d$ for all these cases (during present times) remaining
small compared to 1 au.

\vspace*{-5mm}

\section{Conclusion}

The necessary (but not sufficient) condition to create meteor showers on
Earth as an immediate result of particles ejected from high $e$ sungrazers is
that their orbits lie in a favourable range in $\omega$ thereby enabling the
ascending or descending node to closely approach Earth's orbit. The
forthcoming sungrazing comet C/2012 S1 (ISON) has $\omega \sim 346$\degr. Although this is unusually (for
sungrazers) very close to the ideal condition of $r_d \sim$ 1 au, which would
occur if $\omega \sim$ 348\degr, the descending node nevertheless does not
extend to the Earth's orbit ($r_d \sim$ 0.76 au when $\omega \sim
346$\degr). Even quite high ejection velocities do not bring meteoroids to
intersect the Earth's orbit ($r_d + dr_d \sim$ 0.91 au for 1 km s$^{-1}$
ejection). This implies the absence of strong meteor activity
from this comet.

Compiled observational records of sungrazers (Marsden \& Williams 2008)
reveal only Marsden family comets with $\omega$ lying in this favourable range. The other
sungrazing families have $\omega$ far from this small range during present
epochs and their nodes cannot reach near Earth. This explains why we hardly
see any prominent meteor activity from the frequently observed sungrazers of
different groups.

Surprisingly out of all observed sungrazing family members, none of them have
their orbital elements such that small meteoroid ejection velocities ($\sim
1-100$ m s$^{-1}$) lead to meteor phenomena on Earth (even the Marsden family
typically requiring some hundreds of m s$^{-1}$). It would be interesting to
repeat these calculations for the sungrazers which are going to visit us in
future and check whether any of them have an apt combination of orbital
elements so as to become an exception from this general trend so far.

Furthermore, calculations along these lines can help for forecasting
potential meteor showers on Venus especially because Venus is closer to the
sun compared to Earth (see small nodal distances in Table \ref{orbels},
particularly C/2012 S1 having $r_d$ close to the venusian semi-major axis of
0.72 au). Hence much smaller ejection velocities could induce sufficient
nodal dispersion in meteoroids to reach near the orbit of Venus. This idea
gives much scope for future work using similar techniques. 

 \vspace*{-3mm}

\section*{Acknowledgements}
We appreciate the referee's thoughtful and helpful review. Research at
Armagh Observatory is funded by the Northern Ireland Department of Culture,
Arts and Leisure.

\end{document}